# Bending, breaking, and reconnecting of the electrical double layers at heterogeneous electrodes

Qian Ai[1,2], Lalith Krishna Samanth Bonagiri[2,3], Kaustubh S. Panse[1,2], Jaehyeon Kim[1,2], Shan Zhou[1,2] and Yingjie Zhang[1,2,4]*

1. Department of Materials Science and Engineering, University of Illinois, Urbana, Illinois 61801, United States

2. Materials Research Laboratory, University of Illinois, Urbana, Illinois 61801, United States

3. Department of Mechanical Science and Engineering, University of Illinois, Urbana, Illinois 61801, United States

4. Beckman Institute for Advanced Science and Technology, University of Illinois, Urbana, Illinois 61801, United States

*Correspondence to: yjz@illinois.edu

**Abstract:** In electrochemical systems, the structure of electrical double layers (EDLs) near electrode surfaces is crucial for energy conversion and storage functions. While the electrodes in real-world systems are usually heterogeneous, to date the investigation of EDLs is mainly limited to flat model solid surfaces. To bridge this gap, here we image the EDL structure of an ionic liquid-based electrolyte at a heterogeneous graphite electrode using our recently developed electrochemical 3D atomic force microscopy. These interfaces feature the formation of thin, nanoscale adlayer/cluster domains that closely mimic the early-stage solid-electrolyte interphases in many battery systems. We observe multiple discrete layers in the EDL near the flat electrode, which restructures at the heterogeneous interphase sites. Depending on the local size of the interphase clusters, the EDLs exhibit bending, breaking, and/or reconnecting behaviors, likely due to the combined steric and long-range interaction effects. These results shed light on the fundamental structure and reconfiguration mechanism of EDLs at heterogeneous interfaces.

## Introduction

EDLs exist ubiquitously at the interface between solid electrodes and liquid electrolytes, and play critical roles in virtually all types of electrochemical processes. In supercapacitors, for example, EDLs are responsible for the capacitive charging and energy storage capabilities[1]; in rechargeable batteries, EDLs serve as the precursors for solid-electrolyte interphase (SEI) growth, metal electrodeposition, as well as ion intercalation, thus strongly regulating the energy density and safety of the batteries[2–4]; in various electrocatalytic processes, EDLs are key parts of the interfacial "microenvironment" that modulate the reaction kinetics and product selectivity[5–7].

Despite its paramount significance, the structure and underlying principles of EDL remain elusive. The most widely used classical model, Gouy-Chapman-Stern (GCS) theory, has predicted an EDL structure that consists of a discrete Stern layer and a diffuse Gouy-Chapman layer[8]. While GCS model has shown rough agreements with the EDL capacitance of dilute aqueous solutions (measured by impedance spectroscopy)[9], to date a direct proof from spatially resolved experimental measurements has been absent. Rather, both real-space imaging through 3D atomic force microscopy (3D-AFM) and reciprocal space characterization via X-ray scattering have identified multiple discrete layers of molecular/charged species at the interface between solid single crystals and many different types of electrolytes (dilute aqueous solutions, ionic liquids, water-in-salt electrolytes, etc.)[10–18]. In recent years, it becomes increasingly clear that such multiple discrete layers of liquid are nearly ubiquitously present at solid-liquid interfaces, as a result of the entropic effects inherent to the liquid state[19–22]. Therefore, the molecular density, charge density, and electrostatic potential distribution in the EDL almost always exhibit damped oscillation profiles, instead of the monotonic decays predicted by GCS theory.

While the past decade has seen intensive studies of the EDL structure at flat, crystalline electrode surfaces, real-world electrochemical systems are inevitably heterogeneous. For example, supercapacitors oftentimes use micro/nano-porous electrodes to maximize the surface area[1], alkali-ion battery electrodes tend to be in the form of microparticles to facilitate ion intercalation[23], while electrocatalytic reactions usually utilize nano- or single-atom catalysts to boost the catalytic activity[24,25]. Therefore, knowledge of the EDL structure at heterogeneous interfaces is crucial for the design of electrochemical systems. However, to date heterogeneous EDLs have been a long-standing mystery, with only a few preliminary investigations reporting layers roughly conforming to the solid surface morphology, with limited spatial resolution[14,26,27].

In our previous work, we performed electrochemical 3D-AFM (EC-3D-AFM) imaging of the EDL of an ionic liquid, 1-ethyl-3-methylimidazolium bis(trifluoromethanesulfonyl)imide (EMIM-TFSI), at highly oriented pyrolytic graphite (HOPG) electrode[10–12]. We observed multiple discrete layers, with only the first EDL strongly dependent on the electrode potential. In recent years, due to concerns on battery safety, ionic liquids, with high stability and low volatility and flammability, have gained strong interest for applications in many emerging battery technologies[28–36]. Inspired by these advances, here we study the EDL structure of lithium bis(trifluoromethanesulfonyl)imide (LiTFSI) in EMIM-TFSI at HOPG electrode. Similar to the realistic battery interfaces, in our model system, interphase structures grow at negative electrode potentials, resulting in a heterogeneous electrode–electrolyte interface. We image these interfaces using EC-3D-AFM, and observe rich EDL structural reconfigurations at the heterogeneous sites.

## Results

### Generating controlled heterogeneities

To date the 3D-AFM community has focused on the study of single crystal surfaces, which enable the atomic-resolution imaging of the electrode surface and EDLs[10–13,18,37–40]. The extension of this method to heterogeneous electrode–electrolyte interfaces under operando conditions has been challenging, since electrochemical reactions oftentimes lead to rapid mass transfer and quick generation of gas bubbles, surface cluster/films, etc., all of which significantly perturb/limit the 3D imaging stability and resolution. To generate controlled surface heterogeneities at a slow rate, we resort to ionic liquids, which are highly viscous and inherently sluggish in the mass transfer / ionic transport process. The high stability and slow reaction rate of ionic liquids have also been harnessed in many emerging battery technologies to stabilize the electrode–electrolyte interfaces[28–33]. In our previous work, we have investigated the EDL structure of a widely used ionic liquid, pure EMIM-TFSI. Here we extend the study to a mixture of a low concentration of LiTFSI (0.2 M) in EMIM-TFSI, with the goal of controlled electroreduction to produce surface heterogeneities. While the composition of such heterogeneities closely resembles the typical SEIs in lithium-ion batteries, as we will discuss later, it is not our intention to directly fabricate and test batteries. Rather, we focus on the fundamental structure and reconfiguration mechanism of EDLs at heterogeneous electrodes, which we believe will be pivotal for the further development of a general, beyond-GCS theory to describe the EDL structures at arbitrary electrode–electrolyte interfaces in all types of real-world electrochemical systems.

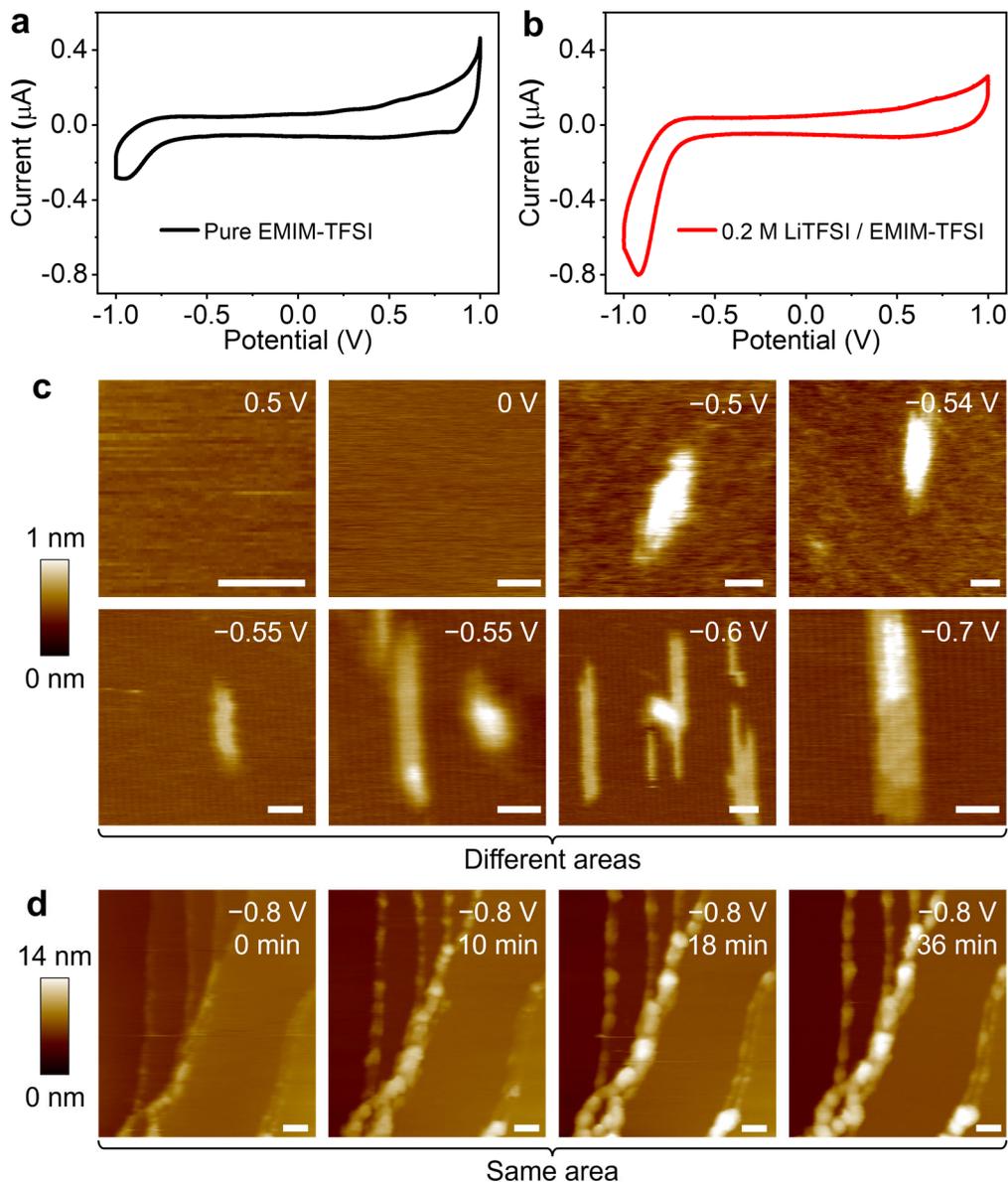

**Fig. 1 | Electrochemical response and surface topography evolution. a** and **b** show the CV of pure EMIM-TFSI and 0.2 M LiTFSI / EMIM-TFSI, respectively, on HOPG electrode (area: 0.64 cm$^2$) with a Pt quasi-reference electrode. **c** Gallery of AC mode EC-AFM topography images (x-y in-plane scan) of the HOPG surface immersed in 0.2 M LiTFSI / EMIM-TFSI, at the corresponding labeled potentials (vs Pt). Each image corresponds to a different area. **d** EC-AFM topography images of a same area of HOPG in 0.2 M LiTFSI / EMIM-TFSI, obtained at a constant potential of −0.8 V vs Pt at different time. Scale bars: 20 nm (all images in (**c**) and (**d**)).

We first examine the overall electrochemical behavior of pure EMIM-TFSI and 0.2 M LiTFSI / EMIM-TFSI systems on an HOPG electrode. As shown in the cyclic voltammetry (CV) results (Fig. 1a), pure EMIM-TFSI exhibits dominantly EDL charging behavior within the 1 V to −1 V potential window (vs Pt). The slight increase in current as the potential reaches close to ±1 V is likely due to the side reactions of impurities, as we previously reported[10]. In the pure EMIM-TFSI / HOPG system, although molecular clusters were sporadically observed, they tend to be tall (>5 nm), weakly adsorbed, and unstable during imaging[10]. To achieve more controlled and stable surface cluster structures, we add 0.2 M LiTFSI to EMIM-TFSI. CV of the mixture electrolyte (Fig. 1b) reveals similar capacitive charging behaviors as pure EMIM-TFSI, although a much larger cathodic current emerges when the potential is more negative than ~ −0.7 V. It is known that $Li^+$ tends to form complexes with $TFSI^-$, mainly in the form of bidentate $[Li(TFSI)_2]^-$ at low salt concentrations[41–43]. On the other hand, the local coordination environment of $EMIM^+$ is expected to be similar in the two electrolytes we have studied. Therefore, the extra current observed in LiTFSI / EMIM-TFSI likely originates from the reduction/decomposition of the $[Li(TFSI)_2]^-$ complexes.

To visualize the structural evolution of the electrode surface, we perform in situ EC-AFM in the x-y surface imaging mode (AC tapping mode) in 0.2 M LiTFSI / EMIM-TFSI. After imaging multiple samples and a series of different areas of each sample, we observe overall three regimes of surface structure: 1) between ~ 0.5 V to −0.5 V, the surface remains clean and flat in most regions, except occasional adsorption of certain clusters at random spots; 2) within ~ −0.5 V to −0.7 V, we observe increased adsorption or nucleation of various nanostructures; 3) more negative than ~ −0.7 V, rapid growth of interphase structures are observed in many areas. Figure 1c, d show representative images at different regimes. The continuous evolution of surface structure at < −0.7 V (Fig. 1d) is consistent with the increase in cathodic current in the CV result (Fig. 1b). In the −0.5 V to −0.7 V regime, the clusters formed on the HOPG surface tend to have an elongated shape with a width of ~ 5–20 nm and a height of ~ 0.5–1.5 nm. With the potential fixed, most of these clusters do not exhibit observable changes over the time scale of one hour (Supplementary Fig. 1). The slow or negligible evolution of these structures is also consistent with the CV result where the redox current does not emerge until the potential is below ~ −0.7 V (Fig. 1b). In the following 3D-AFM investigation of heterogeneous sites, we will focus on the clusters that exhibit negligible changes during the 3D imaging process, so that stable and reproducible 3D images can be obtained.

**3D mapping of EDLs at flat electrode areas**

Before imaging the heterogeneous regions, we first examine the EDL structure at the flat electrode areas within the potential window of 0.5 V to −0.5 V, to both understand the fundamental EDL configuration of the LiTFSI / EMIM-TFSI mixture electrolyte and their potential dependence, and serve as control for further investigation of the more complex cluster sites. Using the same EC-AFM setup, we switch to 3D imaging mode (AC mode, amplitude modulation) to quantify the EDL structure in the chosen flat areas. From the x-z cross section maps, we observe discrete liquid layers on HOPG throughout the measured potential range (Fig. 2a–c), as is typical in ionic liquid systems[44,45].

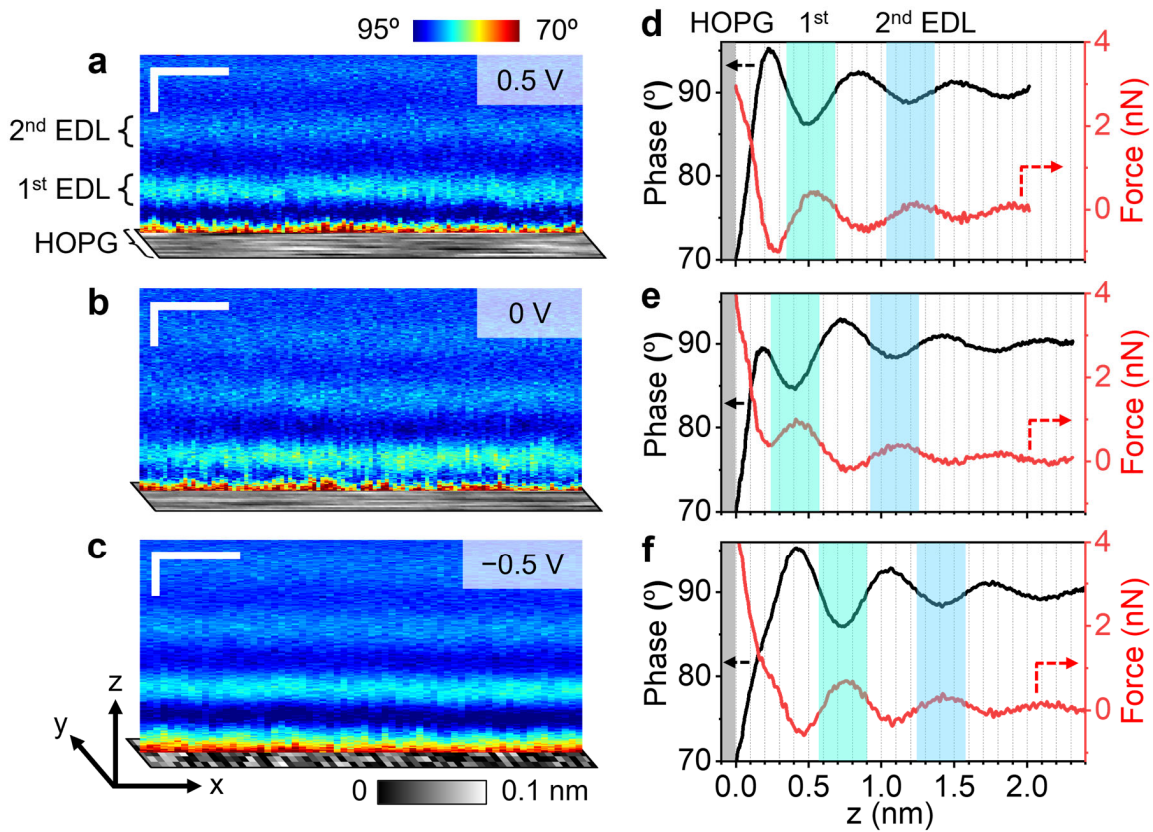

**Fig. 2 | EC-3D-AFM of 0.2 M LiTFSI / EMIM-TFSI on flat HOPG areas. a–c** x-y surface topography and x-z phase maps and **d–f** the corresponding average phase and force vs z curves obtained at 0.5 V, 0 V, and −0.5 V (vs Pt), respectively. The x-y surface topography images (bottom gray-scale maps in (**a–c**)) represent the distribution of z values in the corresponding 3D-AFM maps where the amplitude reaches the setpoint (specified in Supplementary Table 1). In (**d–f**), black and red curves correspond to phase (left y-axis) and force (right y-axis), respectively. Scale bars: 3 nm along x in (**a, b**), and 7 nm along x in (**c**); 0.5 nm along z in (**a–c**). For the x-y surface topography images in (**a–c**), the total y length is 2.3 nm (**a**), 2.5 nm (**b**), and 2.7 nm (**c**), respectively.

Here we choose to use the AFM probe oscillation phase to represent the EDL structure. By averaging the phase maps over the x direction in Fig. 2a–c, we obtain the corresponding phase vs z curves (Fig. 2d–f). Following established protocols, we have reconstructed the force curves and maps from the measured amplitude and phase data (Fig. 2d–f and Supplementary Fig. 2)[46]. Evidently, phase and force curves and maps show nearly the same layered structure. A phase of ~90° corresponds to a freely oscillating probe, i.e., zero probe-sample interaction force, while the local minimums of phase are close to the local maximums of force. In the 3D-AFM community, the exact correlation between the force map and the quantitative EDL structural properties (molecular density, charge density, chemical composition, etc.) is still under debate[12,47–49]. However, it is generally accepted that, larger phase/force oscillation amplitudes (for the same AFM probe) correspond to higher molecular density oscillations in the EDL[19,22,48,50,51]. Therefore, the minimum phase values in each layer are descriptors of the local molecular density — lower minimum phase values, higher molecular density.

We refer the local minimums of phase as "peaks" and assign them as the rough positions of EDLs. Amplitude signals, on the other hand, are less sensitive to the local force variations (Supplementary Fig. 2a–c). Although phase and force signals are nearly equivalent for the purpose of roughly representing the EDL density distribution, phase signal has higher signal-to-noise ratio (SNR) than force, due to the numerical reconstruction process[46] (Fig. 2d–f and Supplementary Fig. 2). Therefore, although the force signal is more direct and quantitative in interpreting the EDL structure, we choose to use phase to represent EDL density variations to maximize the SNR, which will be important for resolving the heterogeneous interfaces later.

Figure 2 reveals that, at all the electrode potentials, the EDL density exhibits damped oscillation behaviors, likely due to the entropic or finite molecular size effects as theoretically predicted[19–22]. As the electrode potential is changed from 0 V to ±0.5 V, the first layer–substrate distance slightly increases, while the interlayer distance of upper layers remains largely constant. This trend is nearly the same as our previously reported EDL structural evolution of pure EMIM-TFSI[10,11], which is consistent with the fact that the EDL charging responses in the CV curves of these two electrolytes are also nearly identical (Fig. 1a, b). The potential-dependent evolution of the EDL structure of 0.2 M LiTFSI / EMIM-TFSI on HOPG was reproduced over multiple separate measurements, with another example shown in Supplementary Fig. 3. As we explained in previous studies, the observed EDL density evolution is likely due to the $EMIM^+$-$TFSI^-$ ion pair reconfiguration in the innermost layer, which is sufficient to screen most of the electrode charge[10,12]. Note that, although $Li^+$-$TFSI^-$ complexes should also exist, the molar ratio of $Li^+$ to $EMIM^+$ is only ~1:19 in the 0.2 M electrolyte we used. Therefore, the overall double layer charging process in the 0.5 V to −0.5 V range should be dominated by $EMIM^+$-$TFSI^-$. This is not contradictory to the higher redox current in 0.2 M LiTFSI / EMIM-TFSI vs pure EMIM-TFSI at potentials below −0.5 V (Fig. 1a, b), as the $Li^+$-$TFSI^-$ complexes (e.g., $[Li(TFSI)_2]^-$), despite having lower concentration, can have more positive reduction potential thus be more easily electroreduced than $EMIM^+$-$TFSI^-$.

### 3D mapping of EDLs at heterogeneous cluster sites

After identifying the EDL structure at flat electrode areas, we proceed to do 3D imaging of the regions where cluster growth has been observed in x-y surface images. Figure 3 summarizes the key features obtained from the 3D images at three different cluster sites at negative potentials. Similar to the HOPG substrate, the clusters also serve as "hard walls" preventing the AFM probe from penetrating through. From the 3D maps, we first extract the surface contour of clusters (Fig. 3a–c), revealing stripe shapes with sub-nm height, similar to those identified in x-y surface imaging in Fig. 1c.

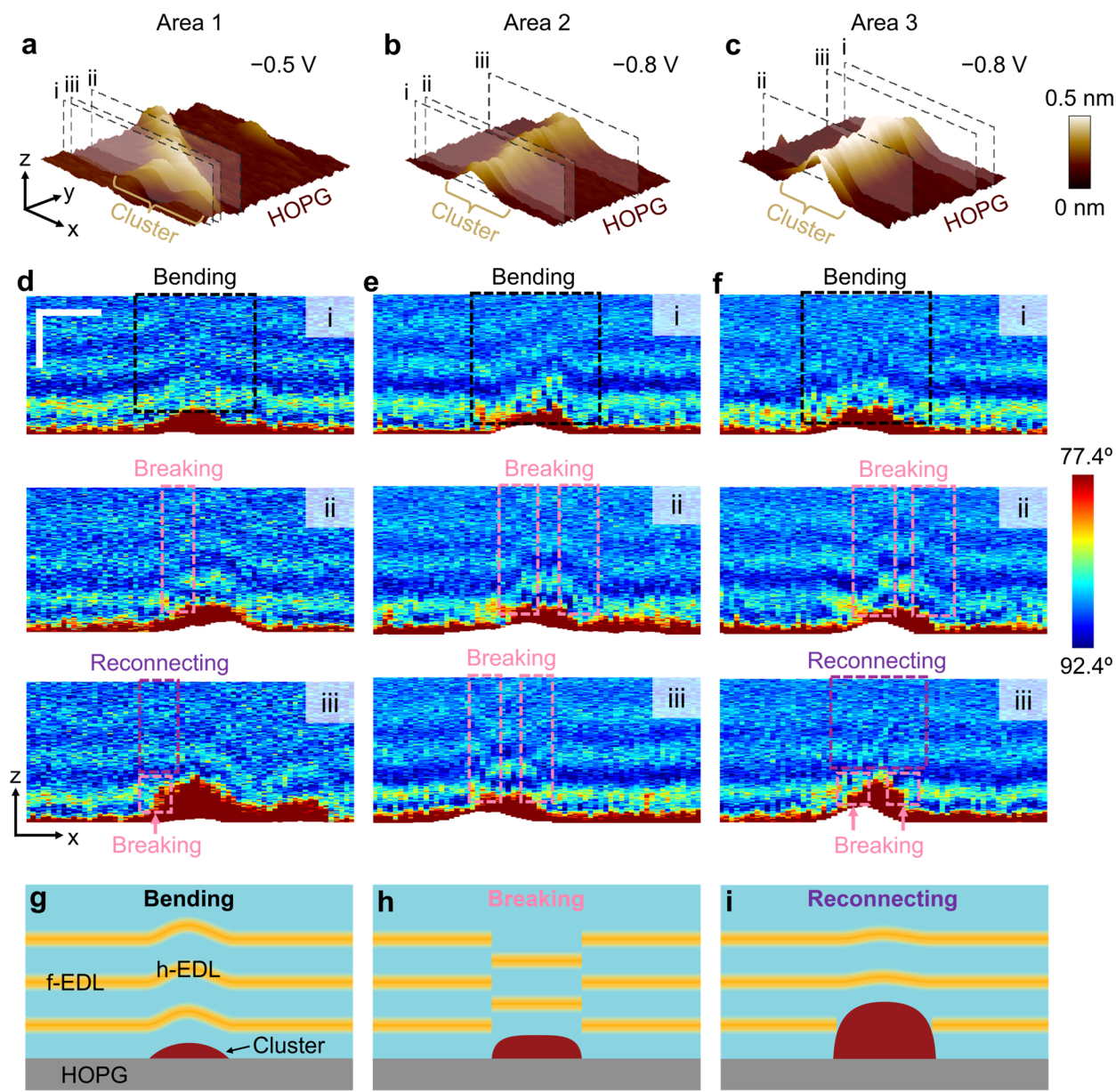

**Fig. 3 | EDL structure at heterogeneous cluster sites. a–c** x-y surface topography of three different areas containing nanoclusters, measured via EC-3D-AFM in 0.2 M LiTFSI / EMIM-TFSI on HOPG electrode. The applied potentials are −0.5 V (**a**) and −0.8 V (**b**, **c**), all vs Pt. The scan size is 100 nm along x (**a–c**), and 100 nm (**a**), 73 nm (**b**) and 75 nm (**c**) along y. **d–f** x-z phase maps measured at the corresponding cross sections marked in (**a–c**). The regions exhibiting obvious "bending", "breaking", and "reconnecting" effects are marked by black, pink, and purple dashed rectangles, respectively. Scale bars: 20 nm along x, and 1 nm along z (same for all the images in (**d–f**)). **g–i** Schematic of the different EDL configurations modulated by the size of the nanoclusters.

The x-z phase maps across different sections of the clusters reveal rich structural features (Fig. 3d–f). When the local cluster is small (i.e., a height within ~3 Å and a width within ~20 nm), the EDLs conform to the height profile of the cluster. That is, the EDLs retain the layered structure, and each layer maintains constant z distance away from the solid surface at the HOPG and cluster areas. Such continuous deformation behavior is marked as "bending" in Fig. 3d(i), e(i), f(i), with a schematic shown in Fig. 3g. In contrast, at cluster sites with a larger size (height and/or width), we observe layer breaking effects (Fig. 3d(ii), e(ii, iii), f(ii)), where the EDLs above the cluster sites abruptly disconnect from the nearby layers above the flat HOPG regions, forming separate liquid layers (Fig. 3h). When layer breaking occurs, the EDLs on top of the cluster area always have z positions roughly in the middle of two adjacent EDL layers above the flat HOPG sites. At sites where the local cluster height is even higher (larger than ~5 Å), the cluster penetrates through the first EDL that extends over the surrounding flat HOPG region, and the EDLs above the cluster exhibit another topological transition marked as "reconnecting" (Fig. 3d(iii), f(iii)). After this transition, the initially broken EDLs are reconnected, with an offset in layer number. That is, the second EDL in the flat HOPG region seamlessly connects to the first EDL above the cluster site, the third EDL above flat HOPG connects to the second EDL on top of the cluster, and so forth (Fig. 3i).

In the following discussions, we refer the local EDL above the flat HOPG area as f-EDL, and those above the cluster sites as h-EDL (heterogeneous EDL) (Fig. 3g).

The full 3D scans of phase maps of the above discussed cluster regions are shown in Supplementary Figs. 4–6. In addition, EC-3D-AFM images of two more cluster regions are summarized in Supplementary Fig. 7. From these extensive sets of 3D maps, we find that the h-EDLs always exhibit one or multiples of the three features: bending, breaking, and reconnecting. These results confirm both the reproducibility and universality of the three types of EDL reconfiguration patterns. Additionally, for the cluster site that fully protrudes out of the first overall EDL, the second overall EDL (containing the first h-EDL) starts to bend again (e.g., in Supplementary Fig. 7d(iv)), as also illustrated in Fig. 3i. Although we have not performed 3D imaging of even taller clusters, it is possible that the "bending-breaking-reconnecting" sequence will repeat at upper EDLs as the cluster becomes higher and higher.

**Quantification of the heterogeneous EDLs**

To gain further insights into the EDL structure and reconfiguration mechanism, we analyze the quantitative phase distribution across the heterogeneous cluster sites. From the three characteristic regions exhibiting bending, breaking, and reconnecting patterns (Fig. 4a–c), we extract the phase vs z curves across the flat HOPG locations and the cluster sites (Fig. 4d–f). At the local HOPG sites away from the clusters, we observe nearly the same phase oscillation behaviors (including the quantitative values) as in Fig. 2d–f, revealing that the clusters have negligible impact on the nearby f-EDLs. Note that the lower SNR in Fig. 4d–f vs Fig. 2d–f is due to the smaller volume of data used for phase averaging. On top of the clusters, we observe vertical offsets in the phase oscillation peaks (Fig. 4d–f), consistent with the spatial features in Fig. 4a–c. At the cluster site where EDL bending is observed, we find that the phase oscillation amplitude at the h-EDL is nearly the same as that of the nearby f-EDL at the corresponding layers, despite the small offset in z positions (Fig. 4d). In contrast, when layer breaking occurs at the cluster site (Fig. 4b), while the h-EDLs still exhibit damped oscillation behaviors, the phase oscillation amplitude value of each local h-EDL layer lies between that of the two adjacent f-EDLs layers (Fig. 4e). As to the cluster site where EDLs reconnect (Fig. 4c), the phase oscillation amplitude of the h-EDL is nearly the same as that of the connected f-EDL (i.e., the first h-EDL connecting the second f-EDL) (Fig. 4f).

In all the local phase vs z curves of h-EDLs, the phase abruptly decreases to below 70° as soon as the cluster surface is reached, similar to those on HOPG surface sites, confirming that the clusters are stiff and behave as hard walls that prevent the AFM probe from penetrating through.

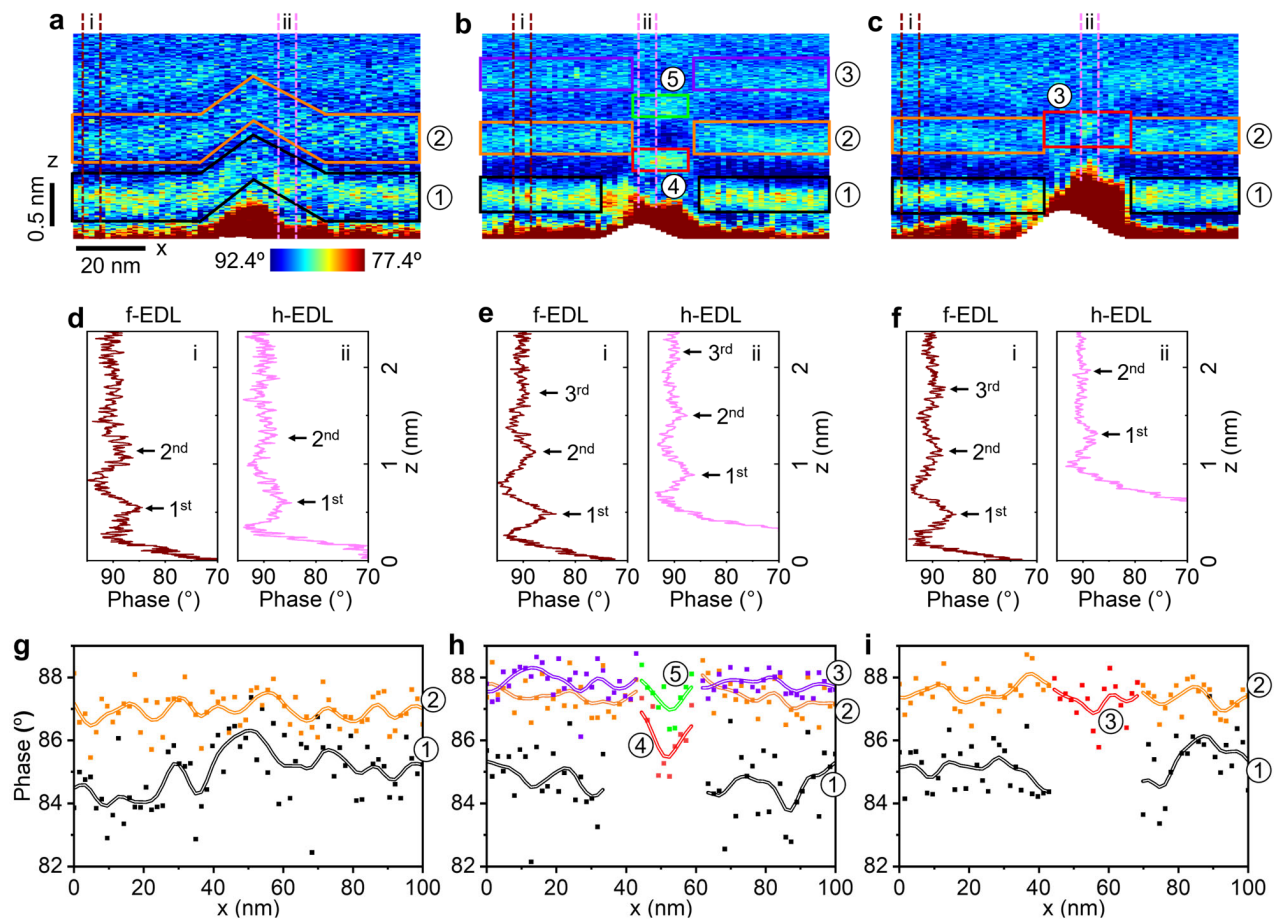

**Fig. 4 | Phase quantification of the heterogeneous EDLs. a–c** x-z phase maps extracted from EC-3D-AFM measurements of 0.2 M LiTFSI / EMIM-TFSI on HOPG electrode, with an electrode potential of −0.5 V (**a**) and −0.8 V (**b–c**) vs Pt, respectively. These maps show the representative EDL bending (**a**), breaking (**b**), and reconnecting (**c**) features. **d–f** Phase vs z curves from the corresponding vertical sections in (**a–c**), respectively, with each f-EDL marked as (i) shown in brown color, while h-EDL as (ii) and pink. The local dips in phase, corresponding to the EDLs, are also labeled by the local layer numbers in (**d–f**). The slightly lower SNR in **d** compared to **e** and **f** is due to the larger z rate (25 Hz vs 10 Hz, Supplementary Table 1). **g–i** Phase vs x plots extracted from the corresponding boxed regions (marked as 1–5) in (**a–c**), respectively. Each scattered point is an average of the 10 lowest phase values in the local z section enclosed in the corresponding box, marked 1–5, colored black, orange, purple, red, and green, respectively. The lines in (**g–i**) are produced by smoothening the corresponding scattered points (MATLAB Gaussian smooth with 8 adjacent points), as visual guides to reveal the phase vs x trend.

We further quantify the in-plane phase evolution across the heterogeneous sites, by plotting the local minimums of phase values at each EDL (average of 10 lowest phase values in a local z section) as a function of x (Fig. 4g–i). Since the free oscillation phase is ~90°, lower minimum phase (below 90°) within each EDL corresponds to higher phase oscillation amplitude. The observed minimum phase vs x plots (Fig. 4g–i) are highly consistent with the examples of phase vs z curves (Fig. 4d–f). Overall, lower EDLs (layers closer to the substrate) exhibit smaller minimum phase, as expected for the damped oscillation patterns. EDL bending, breaking, and reconnecting patterns feature nearly constant minimum phase (Fig. 4g), intermediate phase value (at h-EDL) between two neighboring f-EDLs (above and below the local h-EDL) (Fig. 4h), and constant phase throughout the reconnected EDLs (Fig. 4i), respectively.

The above-discussed trends in the quantitative phase distribution are reproducibly observed in other characteristic cluster sites (Supplementary Fig. 8).

**Discussion**

The observed local EDL reconfiguration across heterogeneous electrode sites is highly intriguing. At the flat HOPG surface, the measured damped oscillation force curves (Fig. 2d–f) are similar to previously reported force profiles of pure solvents at crystalline solid surfaces (e.g., water and octane)[19,52]. Such damped oscillations at liquid–wall interfaces originate from the entropic effects, i.e., the finite size of the liquid molecules combined with their random thermal motion[19–22]. Above the clusters, the local EDLs still exhibit such damped oscillations (Fig. 4d–f), although their overall density oscillation amplitudes become smaller in the breaking and reconnecting configurations.

The local EDL reconfiguration behavior is likely not due to pure electrostatic or electric field effects, since we have observed that changes in electrode potential only modulate the first EDL–substrate distance, not the phase/force oscillation amplitudes (Fig. 2d–f). The possibility of different specific adsorption or chemical interactions between EDL–cluster and EDL–HOPG also cannot explain the observations, as such strong local interactions should lead to different phase/force of the h-EDL vs the f-EDL even in the "bending" configuration, and the h-EDL would not change when the cluster size evolves. Likewise, pure liquid–wall steric effect is also insufficient, since this effect by itself would result in the same EDL oscillation amplitudes above the cluster and HOPG sites.

We propose that the observed EDL reconfiguration features are possibly due to combined steric and long-range interaction effects. Previous surface force apparatus measurements of ionic liquids, including EMIM-TFSI, have revealed long-range electrostatic interactions in the scale of ~10 nm[53–55]. Infrared pump-probe spectroscopy studies of ionic liquids near solid surfaces demonstrated even longer correlation length in the scale of tens of nanometers[56,57]. While our AFM force/phase vs z curves reveal rapid decay of the oscillation amplitude in the scale of inter-layer distance (~0.7 nm) (Figs. 2d–f and 4d–f), it is not contradictory to possible long-range interactions that would extend over a much larger z range and may be beyond the AFM force sensitivity. We expect this long-range interaction to exist not only along z, but also x and y directions. Therefore, the local h-EDLs may tend to retain the same/similar structure and density oscillation as the nearby f-EDLs due to the long-range correlation (Fig. 4). At the same time, steric effects from the cluster surface can lead to vertical position offsets and/or penetration of the h-EDLs. The co-existence and tradeoff between the lateral long-range interaction and vertical steric repulsion are likely responsible for the observed EDL bending, breaking, and reconnecting patterns.

Our observation on the local EDL reconfigurations not only sheds light on the fundamental mechanisms of heterogeneous EDL structure, but also has significant implications for the electrochemical reaction processes. As a model example to illustrate the possible correlation between EDL structure and electrode reaction process, we assume that further charge transfer reaction and cluster growth may occur at more negative electrode potentials in each type of the observed interfacial structure (Fig. 5). We further assume that the reaction is inner sphere, with reactants directly coming from the nearest EDL surrounding the cluster site. At local EDL regions where different phase oscillation amplitude is observed, the molecular density and charge density are likely also different. Variations in charge density distribution can lead to nonuniform electric fields[12], and thus inhomogeneous local $Li^+$ coordination configuration[2,58]. Different $Li^+$ coordination complexes in the innermost EDL will likely induce the deposition and growth of local clusters with different composition[2,4,58–60]. As a result, except for the smallest cluster site where only bending occurs in the EDL, other cluster surfaces will likely feature spatially heterogeneous growth and nonuniform composition, due to the different precursors from the $Li^+$ solvation complexes in the nearest EDL (Fig. 5).

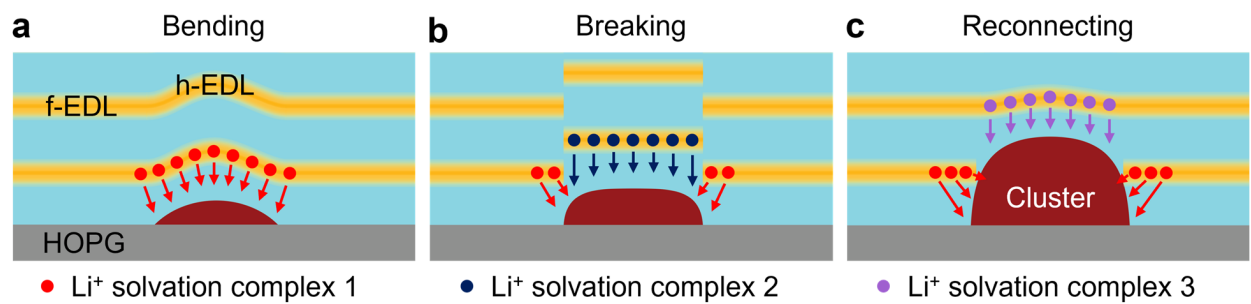

**Fig. 5 | Schematic of possible EDL-modulated electrode reaction processes.** Due to the different EDL configurations, the $Li^+$ solvation complexes are likely uniform in the innermost EDL surrounding the clusters in (**a**), and inhomogeneous in (**b**) and (**c**). As a result, the local composition and growth rate of the clusters can be homogeneous in (**a**) and heterogeneous in (**b**, **c**).

To further quantify the chemical composition of the interfacial structure, we have performed in situ electrochemical Raman spectroscopy under similar conditions as EC-3D-AFM measurements. As shown in Supplementary Fig. 9, we observe characteristic vibrational modes of the HOPG substrate and the electrolyte species, but no additional peaks that can be attributed to the nuclei, likely due to their sparse distribution on the HOPG surface as consistent with Fig. 1c, d. Due to a lack of in situ chemical imaging methods that can detect the grown nanoclusters, we adopt ex situ photo-induced force microscopy (PiFM) to quantify the composition of these structures (Supplementary Note 1). After extended growth under extreme conditions (−1.4 V vs Pt, for a duration of 5 hours), we dry and image the sample in air. From PiFM spectra, we identify a mixture of different species, possibly including $Li_2(NSO_2CF_3)$, $LiSO_2CF_3$, $Li_2S_2O_4$, $(CF_3SO_2)NH$, $LiN(SO_2F)_2$, $Li_2NSO_2F$, $LiSO_2F$, $Li_2SO_3$, $R_1–CH=CH–R_2$, and $NH_2–CH=CH–NH–R$, where R,

R1, and R2 represent alkyl hydrocarbon chains (Supplementary Fig. 10 and Supplementary Table 2.). Although the growth and measurement conditions for PiFM are different from that of EC-3D-AFM, we expect that the same or similar species likely also exist in the in situ observed clusters. The composition of these grown clusters exhibits strong spatial heterogeneity, consistent with the predictions in Fig. 5.

Although we have only investigated one model electrochemical system, we expect the heterogeneous EDL reconfiguration and its correlation to electrochemical reaction kinetics are highly applicable to many practically relevant electrode–electrolyte interfaces. In fact, the cluster compositions and their spatial heterogeneity identified by PiFM are similar to those of the SEIs observed in many lithium-ion battery systems[4,61,62]. SEIs at battery anodes are generally believed to be highly inhomogeneous, although the underlying reason is still largely unknown. Here we propose that the heterogeneity in the local EDLs near the initially nucleated SEIs likely leads to further nonuniform SEI growth, similar to the mechanisms shown in Fig. 5. While the exact correlation between EDL and realistic electrochemical energy conversion and storage functions still cannot be conclusively determined at this stage, our study has already provided significant insights into this "holy grail" problem, and opened a door for a series of further investigations.

## Methods

**Sample and AFM Probe Preparation.** LiTFSI was purchased from Sigma-Aldrich (99.95% purity), and EMIM-TFSI was acquired from Iolitec (99.9% purity). EMIM-TFSI was vacuum annealed for at least 18 hours at 105 ºC before preparing the solution. HOPG (ZYB grade) was obtained from Bruker, and freshly cleaved for use as the working electrode in the AFM electrochemical cell. The surface area of the HOPG exposed to the electrolyte was ~0.64 cm$^2$. Additionally, Pt ring and/or wire were used as the quasi-reference electrode and/or counter electrode. PPP-NCHAuD probes (NanoAndMore) were used for the AC mode EC-3D-AFM imaging. As mentioned in our previous publications[10–12,18], we performed comprehensive calibration of the AFM probes via thermal tune and cantilever tune to determine the spring constants and other probe parameters, as summarized in Supplementary Table 3. Before carrying out the EC-3D-AFM measurements, the probes were cleaned by soaking in acetone for half an hour, in IPA overnight, and in Milli-Q water for a few hours. Finally, the probes were subjected to UV Ozone treatment for 5 mins to remove organic contaminants.

**EC-3D-AFM Measurements.** EC-3D-AFM was carried out using a Cypher ES Environmental AFM (Asylum Research, Oxford Instruments). All the measurements were performed in a sealed argon environment in an electrochemical cell described in our previous publications[10–12,18]. The detailed imaging parameters used to acquire the results are shown in Supplementary Table 1. Prior to either the AC tapping mode x-y topography mapping or the 3D imaging, the cantilever was first driven at the resonance frequency when it was a few microns above the electrode surface. During EC-3D-AFM mapping, the cantilever amplitude and phase were recorded as a function of x, y and z position. We tuned the amplitude set point to ensure that the phase reached at least 70º at the lowest z point at the flat HOPG regions.

**EC-3D-AFM Data Analysis.** We analyzed the EC-3D-AFM data using MATLAB following similar protocols specified in our previous reports[10–12,18]. In the AC mode imaging, the observables are the cantilever amplitude (Amp) and the phase (Phase) as a function of scanner extension (Ext) for each (x, y) pixel. The amplitude values were calibrated using the InvOLS (inverse optical lever sensitivity) parameter (obtained from the thermal tune). The phase values were offset to ensure 90º is reached when the probe is sufficiently far away from the surface (corresponds to z > 2.5 nm) The z values were initially obtained at each (x, y) point using the formula z = − Ext − Amp, and then offset to achieve an average of z = 0 at regions corresponding to the HOPG surface. We set 70° phase value as the substrate position because it corresponds to a high enough force to ensure the probe reaches the substrate, as shown in our previous publications on the comparison of AC vs DC mode 3D-AFM in ionic liquids[10,11]. Due to the sharp decrease of phase as the probe reaches the substrate, setting the substrate phase to other values near 70° only leads to minimum change of the substrate position (Supplementary Fig. 11). At the heterogeneous cluster areas, the overall 3D data were tilt-corrected to ensure an average z=0 at the flat, clean HOPG regions.

Force reconstruction in this work was carried out using Eq. 6a in Payam et al[46]. The input parameters for the force calculation include Amp and Phase vs z as well as the AFM probe metrics including spring constant, quality factor and resonance frequency. All of these probe parameters are summarized in Supplementary Table 3.

**Electrochemical Raman measurements.** The EC-Raman experiments were performed using the same electrochemical cell as EC-3D-AFM measurements. We used a Raman confocal imaging system (Horiba LabRAM HR 3D-capable Raman spectroscopy) with 300 grooves/mm grating. The wavelength of the laser was 633 nm. The original laser power of ~35 mW was reduced to ~3.5 mW by applying an optical filter. The laser was switched on for more than 30 minutes before the spectra collection and was then focused on the electrode surface using a 50× objective. At least one minute after applying a potential step, a Raman spectrum was obtained over three accumulations, with 30 seconds integration time per accumulation. We used a Si wafer for spectral calibration.

**PiFM Measurements.** The sample (nucleated clusters on HOPG substrate) was grown inside an argon filled glovebox (<0.1 ppm of $O_2$ and $H_2O$). A CHI 600E potentiostat was used. The working electrode was HOPG, and the reference electrode was a platinum ring (in the same configuration as EC-3D-AFM and EC-Raman). These clusters were grown by applying an electrode potential of −1.4 V vs Pt for a duration of 5 hours. Then the sample was removed from the glovebox, rinsed thoroughly with IPA, and then blow-dried with nitrogen gas. The PiFM measurements were then performed using a Vista One PiFM-Raman microscope from Molecular Vista. All the PiFM measurements were performed in ambient air. The spectra were acquired using a quantum cascade laser from Block Engineering with a wavenumber range of 752–1908 $cm^{-1}$ and a resolution of 0.5 $cm^{-1}$. The measurements were carried out in the so-called "dynamic mode", using a PPP-NCHR probe from NanoSensors custom coated with Pt-Ir.

**Cyclic Voltammetry.** CV was performed using the AFM electrochemical cell in an argon-filled glovebox (< 0.1 ppm of $O_2$ and $H_2O$). A CHI 600E potentiostat was used to sweep the potential and record the current. The working, reference and counter electrodes were HOPG, platinum ring and platinum wire, respectively. The active surface area of the working electrode was about 0.64 $cm^2$ and the volume of the liquid used in the measurements was ~120 μL.

**Data availability**

All the data shown in this paper are available from the corresponding author upon reasonable request.

## Code availability

The computer codes used to generate the results reported in this study are available from the corresponding author upon reasonable request.

## Acknowledgements

This material is based upon work supported by the Air Force Office of Scientific Research under award number FA9550-22-1-0014. The experiments were performed in part in the Carl R. Woese Institute for Genomic Biology and in the Materials Research Laboratory at the University of Illinois Urbana-Champaign. The authors also acknowledge Dr. Kathy Walsh for help with PiFM measurements.

## Author contributions

Q.A. did data analysis and co-wrote the manuscript. L.B. did data analysis. K.P. carried out EC-3D-AFM and PiFM experiments. J.K. carried out electrochemical Raman experiments. S.Z. contributed to EC-3D-AFM experiments. Y.Z. designed the project, supervised the work, and wrote and revised the manuscript. All authors reviewed the manuscript.

## Competing interests

The authors declare no competing interests.

## Additional information

**Supplementary information** The online version contains supplementary material.